\documentclass[manuscript,nonacm]{acmart}

\usepackage{pdfpages}
\usepackage{subfigure}

\AtBeginDocument{%
  \providecommand\BibTeX{{%
    \normalfont B\kern-0.5em{\scshape i\kern-0.25em b}\kern-0.8em\TeX}}}

\acmConference[KDD 2025]{ACM Conference on Knowledge Discovery and Data Mining}{3--7 August 2025}{Toronto, ON}
\begin{document}

\title[Artificial Intelligence in Environmental Protection]{Artificial Intelligence in Environmental Protection: The Importance of Organizational Context from a Field Study in Wisconsin}


\author{Nicolas Rothbacher}
\authornote{These authors contributed equally to this work.}
\email{nsroth@law.stanford.edu}
\author{Kit T. Rodolfa}
\authornotemark[1]
\orcid{0000-0002-0829-1282}
\email{krodolfa@law.stanford.edu}
\author{Mihir Bhaskar}
\email{mihirb@law.stanford.edu}
\author{Erin Maneri}
\email{emaneri@law.stanford.edu}
\author{Christine Tsang}
\email{ctsang@law.stanford.edu}
\author{Daniel E. Ho}
\orcid{0000-0002-2195-5469}
\email{deho@law.stanford.edu}
\affiliation{%
  \institution{Stanford University}
  \streetaddress{RegLab, 559 Nathan Abbot Way}
  \city{Stanford}
  \state{California}
  \country{USA}
  \postcode{94305}
}

\renewcommand{\shortauthors}{RegLab}

\begin{abstract}
  Advances in Artificial Intelligence (AI) have generated widespread enthusiasm for the potential of AI to support our understanding and protection of the environment. As such tools move from basic research to more consequential settings, such as regulatory enforcement, the human context of how AI is utilized, interpreted, and deployed becomes increasingly critical. 
  Yet little work has systematically examined the role of such \textit{organizational} goals and incentives in deploying AI systems. We report results from a unique case study of a satellite imagery-based AI tool to detect dumping of agricultural waste, with concurrent field trials with the Wisconsin Department of Natural Resources (WDNR) and a non-governmental environmental interest group in which the tool was utilized for field investigations when dumping was presumptively illegal in February-March 2023. Our results are threefold: First, both organizations confirmed a similar level of ground-truth accuracy for the model's detections. Second, they differed, however, in their overall assessment of its usefulness, as WDNR was interested in clear violations of existing law, while the interest  group sought to document environmental risk beyond the scope of existing regulation. Dumping by an unpermitted entity or just before February 1, for instance, were deemed irrelevant by WDNR. Third, while AI tools promise to prioritize allocation of environmental protection resources, they may expose important gaps of existing law. 
\end{abstract}



\received{10 February 2025}

\maketitle

\section{Introduction}
One large cattle farm can produce more manure each year than the waste produced by Houston, Texas, a city of more than 2 million people \cite{mittal2009concentrated}. Improperly handled, this waste can pose considerable threats to the safety of nearby waters and the health of local communities \cite{usepa2003national,bradford2008reuse}, but identifying instances of waste mismanagement can be difficult for regulators. Seeking to reduce this barrier to regulatory enforcement, recent research demonstrates how a computer vision model could make use of near real-time satellite data to detect potentially hazardous manure management practices \cite{land_app_CIKM}. As a promising application of Artificial Intelligence (AI) to environmental sustainability that work is far from alone, with recent years having seen dramatic growth in proposed uses of these tools to better understand and protect our environment \cite{handan2021deep,nishant2020artificial,kar2022can}: Machine Learning (ML) models can help focus limited resources where they can do the most good, whether helping regulators target facilities likely to violate environmental protections \cite{hino2018machine} or conservationists seeking to prioritize areas that will most promote biodiversity \cite{silvestro2022improving}. Advances in AI and edge computing may improve deep sea research and support efforts to promote ocean health \cite{molina2021autonomous,battula2020online}, while new datasets for computational fluid dynamics can improve our understanding of how the climate evolves \cite{chung2023turbulence}. The combination of modern computer vision methods and more timely, higher-quality remote sensing data has been a particularly active avenue for research \cite{handan2021deep}, with applications ranging from predicting wildfire risk \cite{jain2020review,jaafari2019hybrid}, identifying factoring farming operations \cite{handan2019deep,robinson2021mapping,robinson2021temporal,chugg2021enhancing} and estimating their climate impacts \cite{jeong2022artificial}, measuring changes in polar ice composition \cite{mosadegh2022new}, to even helping researchers track and study whale populations from space \cite{kapoor2023deep,hodul2023individual}. While the fast pace of recent research and development efforts has surfaced a wide range of interesting use cases, such systems should not be considered in isolation. In consequential settings, AI models will inherently interact with humans and organizations, meaning that properly evaluating their utility necessitates understanding how they operate as one component of a larger sociotechnical system \cite{amarasinghe2023explainable,schwartz2022towards,ackermann2018deploying,fogliato2022case,cheng2022child,levy2021algoinpublic}. 

\begin{figure*}[t]
\includegraphics[width=0.37\textwidth]{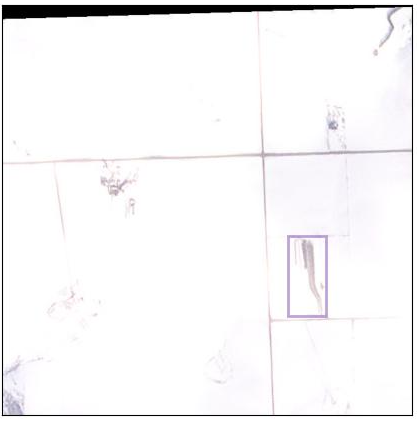} 
\hspace{0.2cm}
\includegraphics[width=0.50\textwidth]{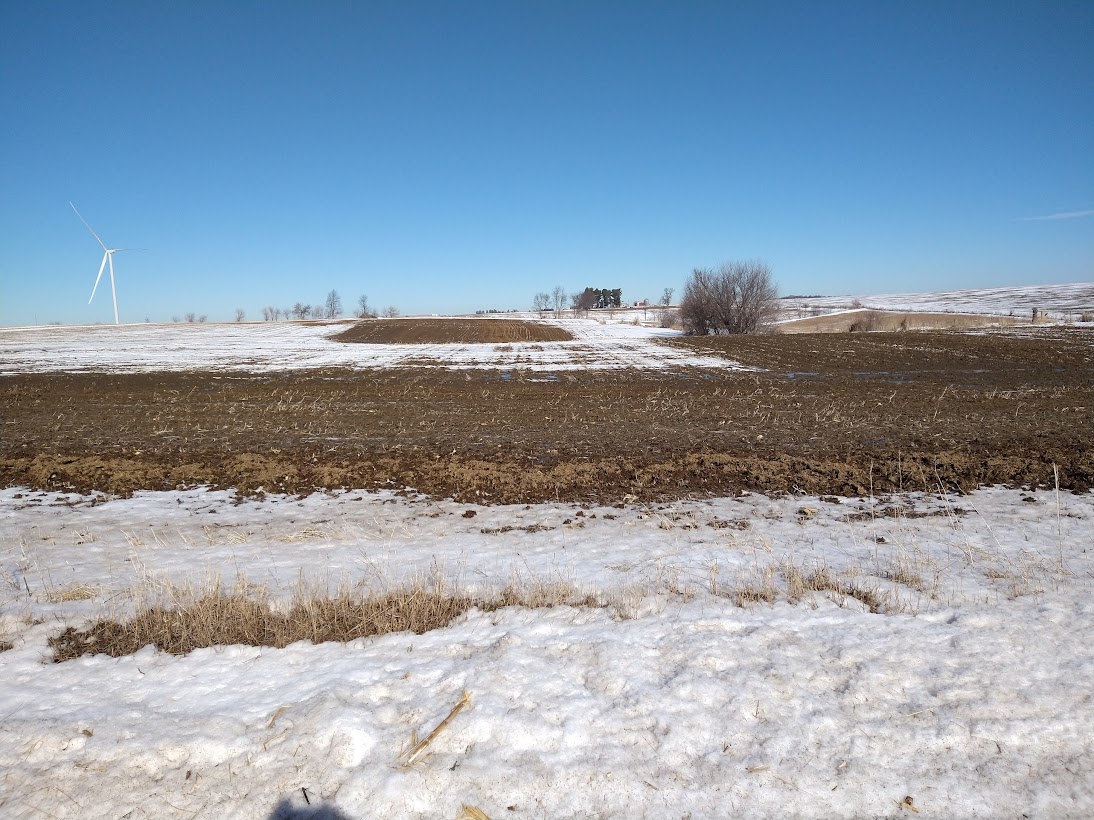} 
\centering
\caption{Example of satellite imagery (left) and ground-verified (right) detection of manure spreading in Grant County, WI.}
  \Description{The left panel shows satellite imagery with a model detection indicated by a purple bounding box. The right panel shows a photo of this same manure spreading event taken from the ground by one of the ELPC verifiers.}
\label{fig:elpc_photo}
\end{figure*}

In prior work \cite{land_app_CIKM} we presented a new approach to land application monitoring using computer vision on high frequency satellite imagery. This model was developed in partnership with the Environmental Law and Policy Center (ELPC) who advised on the nature and appearance of land application and assisted in initial testing and improvement of the model in the winter of 2022. The present study builds on this partnership with ELPC to perform an extensive field trial of the model in the winter of 2023. During the planning of our pilot, our contacts at the Wisconsin Department of Natural Resources (WDNR), the state agency charged with enforcing environmental policy in partnership with the US Environmental Protection Agency (EPA), also became interested in piloting the model in their work. This provided the unique opportunity to study how the model could be deployed in multiple settings.

In the present work, we study the organizational context for utilizing, interpreting, and deploying the model introduced in \cite{land_app_CIKM} through unique field trials with WDNR and ELPC, occurring in February-March 2023. The environmental and health risks in this setting are significant: 
In 1998, the US Government Accountability Office estimated that American Concentrated Animal Feeding Operations (CAFOs) produced more than 13 times the annual amount of human waste across the United States \cite{burkholder2007impacts}. 
Although animal manure can be valuable as a fertilizer for crops, excess manure becomes an environmental hazard and its overuse can harm environmental and human health \cite{bradford2008reuse}. 
Of particular concern is the high concentration of phosphorus and nitrogen in manure which, when washed off into waterways, can cause mass die offs and dead zones for aquatic animals through eutrophication\footnote{Eutrophication is the overgrowth of plant life in water that deprives other aquatic life of oxygen.} \cite{mallin2003industrialized}. Manure application may also contaminate nearby surface and groundwater with fecal matter \cite{mallin2015industrial}, heavy metals \cite{liu2015arsenic, nachman2005arsenic}, antibiotics \cite{campagnolo2002antimicrobial}, microbial pathogens \cite{gerba2005sources}, and endocrine disrupting hormones \cite{hanselman2003manure, raman2004estrogen}.

Land application of manure onto farmland in the winter months has been a particular target of environmental policy \cite{liu2018review, srinivasan2006manure, bihn2019oversight} because it carries higher risks: frozen or wet ground coupled with the lack of growing crops increases the runoff of manure and creates higher risk of ground or surface water contamination \cite{klausner, lewis2009winter, srinivasan2006manure, williams2011manure}. In Wisconsin, winter land application by CAFOs is strictly regulated by WDNR. In February and March, liquid waste is generally illegal to apply to any field, and solid waste is not permitted on any snow-covered or frozen field (\S~NR 243.14(6)(c)), except in rare cases of ``emergency applications'' approved by WDNR. Such prohibitions, however, do not apply to months outside of Feburary and March or to Animal Feeding Operations (AFOs) with less 1,000 animal units. This results in a hard cutoff where smaller facilities are in some cases allowed to apply manure while farms that may have just a few additional animals are prohibited from applying during February and March. While permitted CAFOs do tend to produce more manure and larger environmental impacts, the collective production of all the smaller farms may require more scrutiny as the total amount of manure released into the ecosystem is what produces adverse impacts on the environment. Indeed, facilities that fall short of the 1000 animal unit threshold are the most common type of animal farming operations in Wisconsin: an analysis of satellite imagery in 9 counties in Wisconsin found that only 3 percent of observed animal feeding operations had permits as CAFOs \cite{ewg2022wicafos}. 


We report results from field trials with both WDNR and the ELPC, which conducted concurrent and independent pilots of the real-time monitoring system for land application introduced by \cite{land_app_CIKM} (Figure~\ref{fig:elpc_photo} shows an example of a detection viewed in both satellite imagery and from the ground). These field trials were a significant undertaking: ELPC's verifiers alone drove approximately 4,300 miles across the state, spending 175 hours visiting the sites of model detections. While both pilots sought to provide ground-truth validation of the model's detections of manure spreading, WDNR, as the state's regulatory authority, was also positioned to assess---and was particularly interested in---whether each detected application was in violation of environmental regulations. Running field trials with both organizations at the same time allowed us to explore how their varying goals and perspectives informed their assessment of the system, a question that has received little attention in the literature on human-computer interactions (see the Appendix for a detailed review). Despite finding very similar rates of verified detections in both trials, WDNR saw relatively few of these translate into clear regulatory violations that were within their purview, limiting their enthusiasm for a longer-term deployment of the tool. By contrast, ELPC's interest in measuring the overall extent of winter spreading and its potential environmental impacts (including from events that are compliant under current regulations) led to a more positive outlook on the pilot.

Taken together, this work makes three key contributions to the literature. 
First, we present a boots-on-the-ground field trial to validate a tool that allows for near-real-time remote sensing of environmental harms, demonstrating that the model provides substantial gains in efficiency relative to manual review of satellite imagery and in coverage relative to current complaint-driven processes. 
Second, the concurrent field trials with both a government regulator and advocacy organization provide a novel opportunity to assess  the role of organizational context on the utilization, interpretion, and deployment of machine learning systems. 
Third, although our study confirms considerable winter manure application, much of this dumping escapes existing regulations. This stems from temporal thresholds (dumping is legal on Jan. 31, but illegal Feb. 1) and size definitions (dumping is legal by an AFO with 999 animal units, but illegal by a CAFO of 1,001 animal units) in environmental law. In the present study, 82\% of the 
dumping activity confirmed by WDNR fell between these regulatory cracks. The hard cutoffs under current policy create leeway for manure dumping onto snow covered ground, that we detected and confirmed, a practice that is targeted specifically for restriction under state law but is still being allowed due to the broad exceptions. These results highlight the need for policymakers to carefully consider how policy design choices can serve to promote or undermine both compliance and the underlying societal goals.

\section{Methodology}
This study focuses on the implementation and use of a machine learning decision support system to understand how two different organizations focused on the same issue would utilize the tool and react to its recommendations. Prior to the trial, we collected data, trained the model, implemented a real-time prediction delivery system, and worked with partners to identify and coordinate staff to investigate predictions. Throughout the study, we worked closely with those partners to update procedures to ensure the system was working efficiently with the verifiers and collected responses from them. Finally, both during and after the study period, we interviewed verifiers individually and in groups to understand how they used the system, reacted to its recommendations, and would improve or use the system in the future.

\begin{figure*}[t]
\includegraphics[width=0.30\textwidth]{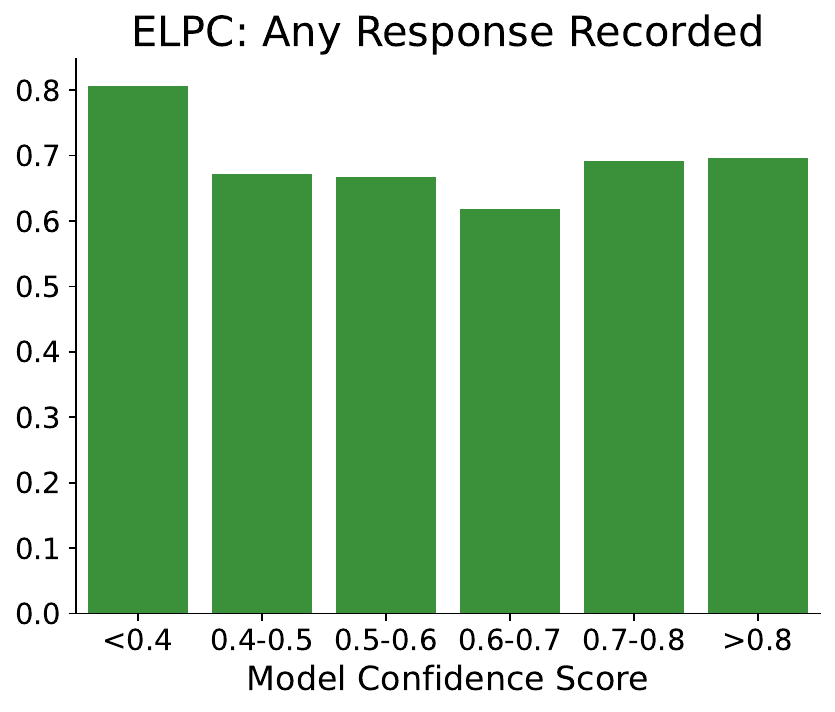} 
\hspace{0.2cm}
\includegraphics[width=0.30\textwidth]{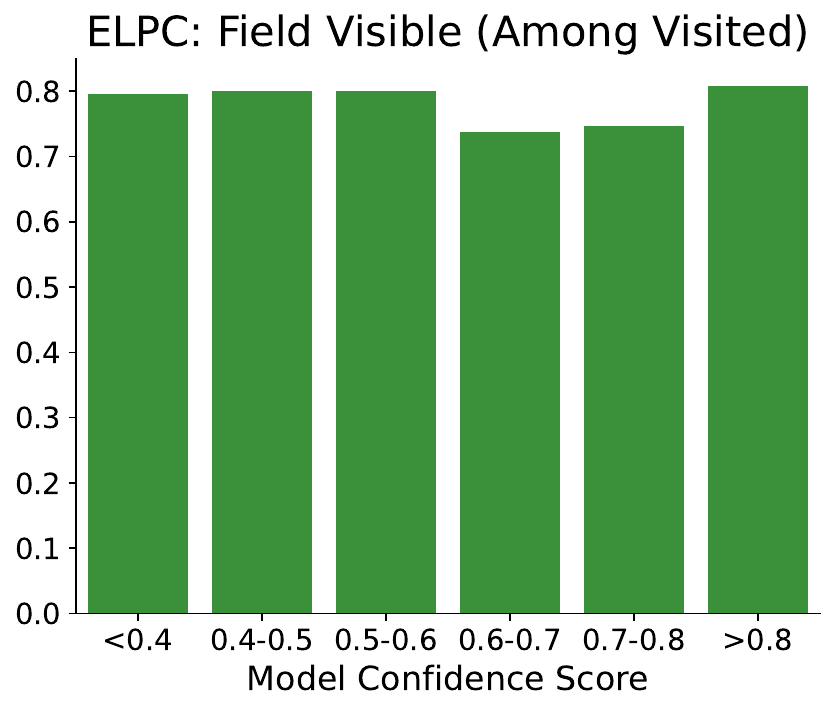} 
\hspace{0.2cm}
\includegraphics[width=0.30\textwidth]{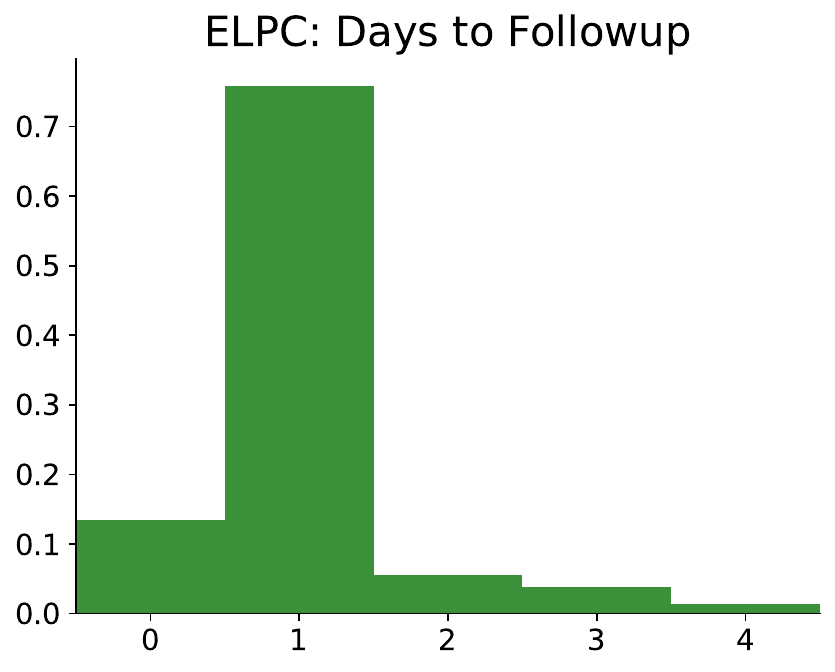} 
\centering
\caption{Process metrics from the ELPC trial. (A) Follow-up rate by model score. (B) Among detections visited, ELPC verifiers were able to see the location from public roads at similar rates across model scores. (C) Distribution of days to follow-up.}
\label{fig:elpc_process_metrics}
\end{figure*}

\subsection{Data Sources and Model Training} 
The machine learning system tested here builds on the modelling approach first described by Chugg and Rothbacher, et al. \cite{land_app_CIKM} and makes use of two primary data sources: near-daily visual spectrum satellite imagery, and information about known CAFO locations throughout Wisconsin based on permit data from WDNR. We prioritized the collection of training data, real-time model runs, and detections around the known CAFO locations to send to participants as part of the study. 

Satellite imagery was collected for both training and running the model in real-time for the study. All input satellite data was collected from a subscription with PlanetLabs that provided 3m/pixel resolution 3-band visual spectrum imagery\footnote{For this imagery, the three bands are red, green and blue. The imagery is also orthorectified, color corrected, sharpened and corrected for sun angle \cite{planet}}. We use training data collected by \cite{land_app_CIKM} which consisted of 2 sqkm images centered on permitted CAFO facilities, which were hand labeled by at least two human labelers to place bounding boxes around suspected land application events. The authors applied this procedure to daily imagery from November, December, January, February, and March of the winters of 2018-19, 2019-20, and 2020-21. This resulted in a total of 1,061 images containing application and 1,813 total instances of application across 96 CAFO locations in the prior study \cite{land_app_CIKM}. 

Data for model inference runs during the study was collected from the previous day to the model run using the same PlanetLabs data product as the training data. These model runs used larger areas of interest (6 sqkm) around permitted CAFO locations and satellite sites\footnote{CAFO satellite sites are smaller facilities identified by permittees often used for waste storage and disposal or smaller animal holding facilities.}. These locations were provided by WDNR. We also included some locations outside of this set in Door County to cover some known areas of manure spreading away from permitted facilities.

In addition to the locations of CAFO permits and their satellite sites, WDNR provided a geospatial shapefile of permitted fields for manure spreading. These fields are reported in CAFO NMPs for compliance with state regulations on the types of fields and setback distances from risky areas like wells and roadways where manure can be legally applied \cite{wi_guidelines}, and were used to filter model detections for follow-up by WDNR. ELPC followed up on all detections, including those not on NMP fields. 

The model architecture and fitted weights, as we previously reported in \cite{land_app_CIKM}, were produced by fine-tuning the You Only Look Once (YOLO) pre-trained object detection model (version 5) \cite{yolov5} for the land application task using the training set of labeled land application detection images. This model produces a bounding box and confidence score for each detected instance of land application present in a satellite image. 
For the purposes of the field trial, we developed a production pipeline\footnote{Code for this pipeline will be made available on a public repository upon publication of this work.} using python and postgresql to wrap the trained model to run inference on new imagery and route detections to our partner organizations. In production mode, the model runs inference regularly over recently captured imagery prioritized by the locations of permitted CAFOs and satellites and produces instance detections and confidence scores. Here, inference was focused on a 6km box around these known CAFO locations, based on conversations with our partners and reflecting an expectation that a facility would be unlikely to haul manure a long distance for spreading, particularly with winter road conditions.


\subsection{Study Design}
We designed what were essentially two separate studies for the two organizations we worked with, each with their own participants, detection prioritization, detection delivery, and response collection protocols. Both relied on the same satellite imagery and model runs but limited their delivery of detections to participants on different criteria. We aimed to provide detections twice a week in the months of February and March, but due to weather and imagery constraints, imagery could not be captured at all desired locations during every model run in which case, any locations with satellite imagery were run through the model and detections were provided for each organization. Prior to the beginning of the study in earnest, multiple trial runs were undertaken with the two groups in December and January. These involved sending imagery and detections from these earlier dates in order to test and evaluate detection distribution, prioritizing, and validating procedures.

\subsubsection{Wisconsin Department of Natural Resources}
We worked with the State of Wisconsin's Department of Natural Resources (WDNR) to pilot the model with the team that is devoted to reviewing CAFO applications and working with permittees to comply with permit requirements. Employees at the central office routed model detections to regional CAFO specialists who work closely with certain permittees to review for violations as they would with citizen-reported complaints of spreading. For WDNR, detections were filtered to those that were both above model confidence 0.5 and were geolocated to be on a field permitted to be used for land application by a CAFO. These detections were then screened visually by the central WDNR office for those determined to be more likely than not actual manure application. This effort sought to eliminate obvious false positives such as vegetation and buildings. Unlike the ELPC verifiers, the WDNR staff had the authority to directly follow-up with the regulated farms. Upon receiving detections, these specialists worked with CAFO permittees or on their own to validate the detections and determine whether the detected application was in violation of Wisconsin law, through a combination of site visits and communication with CAFOs, smaller farms, and field owners. Data on each detection was collected by the WDNR central office and distributed to us after quality control. Participants were asked to report the date of their followup, steps that they took to determine existence and regulatory compliance of detected manure application, and reasons why they made their determination. We also conducted individual post-trial period interviews with three of the participating specialists to discuss their use of the system, experiences they had following up on detections, and the reactions of permittees to their inquiries. Following the conclusion of the pilot, we held a debrief session with the central office staff and conducted semi-structured interviews with three of the regional CAFO specialists (the script for these interviews is included in the Supplementary Methods).

\subsubsection{Environmental Law and Policy Center}
We also worked with the Enviromental Law and Policy Center (ELPC), a legal and political environmental advocacy group that operates in the Midwest of the United States to identify and recruit citizen advocates in select counties across Wisconsin. 15 people were selected to participate based on their willingness and ability to drive from detection to detection and report information about the ones they were selected to verify. Model detections were filtered to those that were within 25km of a participant. For each model run, the detections with the 5 highest confidence scores within that 25km were selected for each participant. These detections were then given to ELPC, who coordinated the dispatching of participants to observe detected application instances. Participants were distributed satellite images of the application with a detection bounding box, a side-by-side comparison image taken from Google Maps, the date of the satellite image capture, and the coordinates of the detection site for navigation to the location. Participants were asked to observe in-person all detections routed to them and respond to a Google survey form. Because ELPC's verifiers could only make their observations from public roads, this survey asked then to report whether they could identify the location of the detection and, if so, their assessment of the presence of manure, as well as their level of certainty. The full survey questionnaire can be found in the Supplementary Methods. This survey data was collected into a single database of responses. In addition to the observation data, we interviewed participants in groups to solicit feedback and understand how participants were using and reacting to the detections. We held two meetings with participants two weeks into the study period and another two meetings after the conclusion of the study period.

Although a pre-screening process was used by WDNR staff to only route the most likely manure detections for follow-up by field CAFO specialists, the ELPC trial sought to develop on-the-ground verification of all model detections that were sent. The group's verifiers drove approximately 4,300 miles across the state, spending 175 hours visiting the sites of model detections. Responses were recorded from a similar proportion of detections across the model's confidence score, suggesting our trial results are representative across the range of scores sent (Figure~\ref{fig:elpc_process_metrics}A). Likewise, among detection locations visited, verifiers were able to identify the location from public roads 77\% of the time, with a consistent visibility rate across the model score distribution (Figure~\ref{fig:elpc_process_metrics}B). These follow-up visits were timely as well (Figure~\ref{fig:elpc_process_metrics}C), reducing the risk of manure applications being obscured by thaw or covered by additional snow: among visited sites, 90\% were within 1 day of the detection being sent to ELPC and none took more than 4 days for follow-up.

\section{Results}
\subsection{Verifiying Dumping Detections}


\begin{figure*}[t]
\subfigure[All detections sent to WDNR]{
\includegraphics[width=0.3\textwidth]{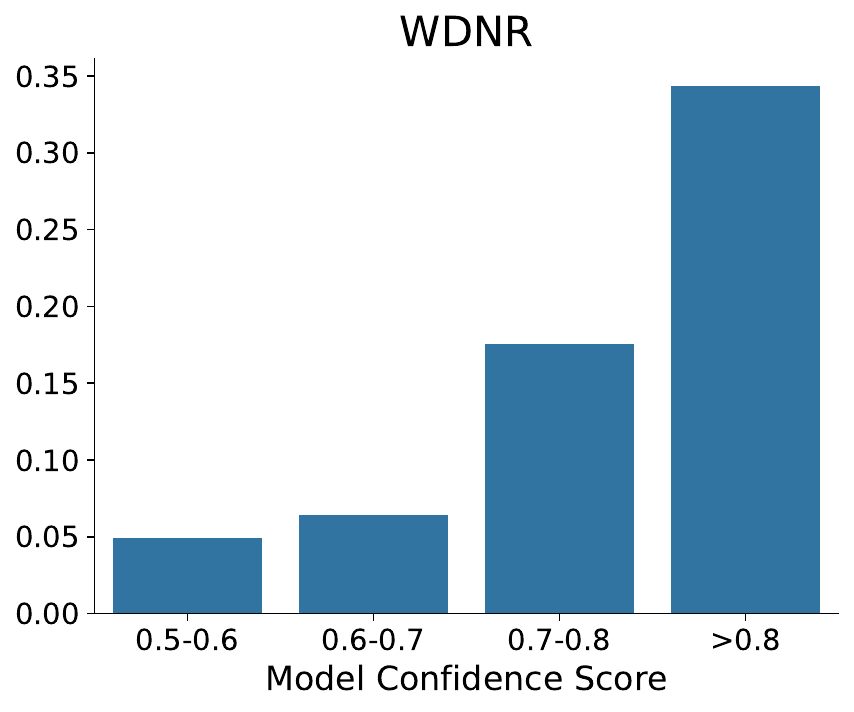}}
\hspace{0.1cm}
\subfigure[WDNR detections after initial review]{
\includegraphics[width=0.3\textwidth]{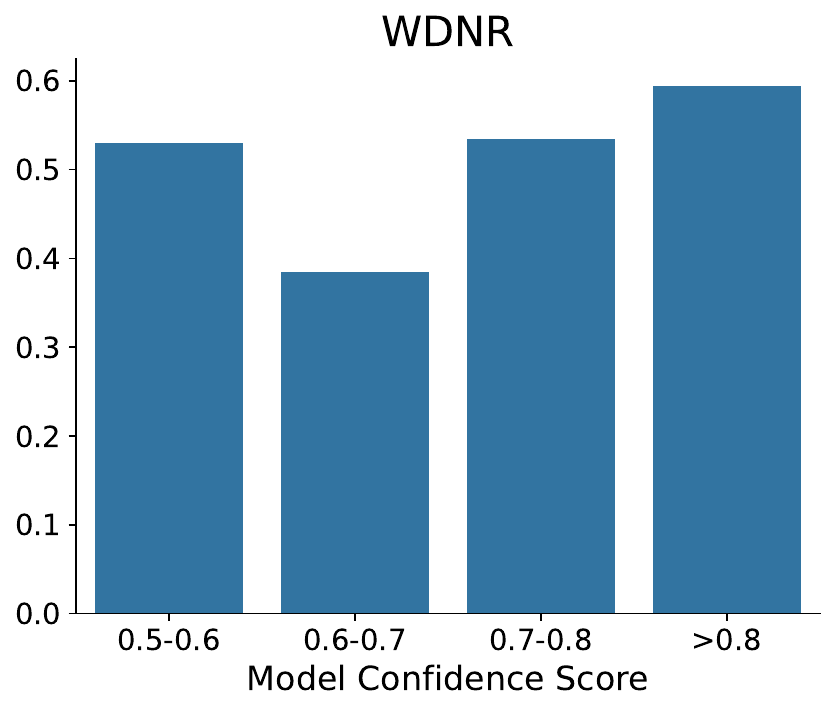}}
\hspace{0.1cm}
\subfigure[All detections sent to ELPC]{\includegraphics[width=0.3\textwidth]{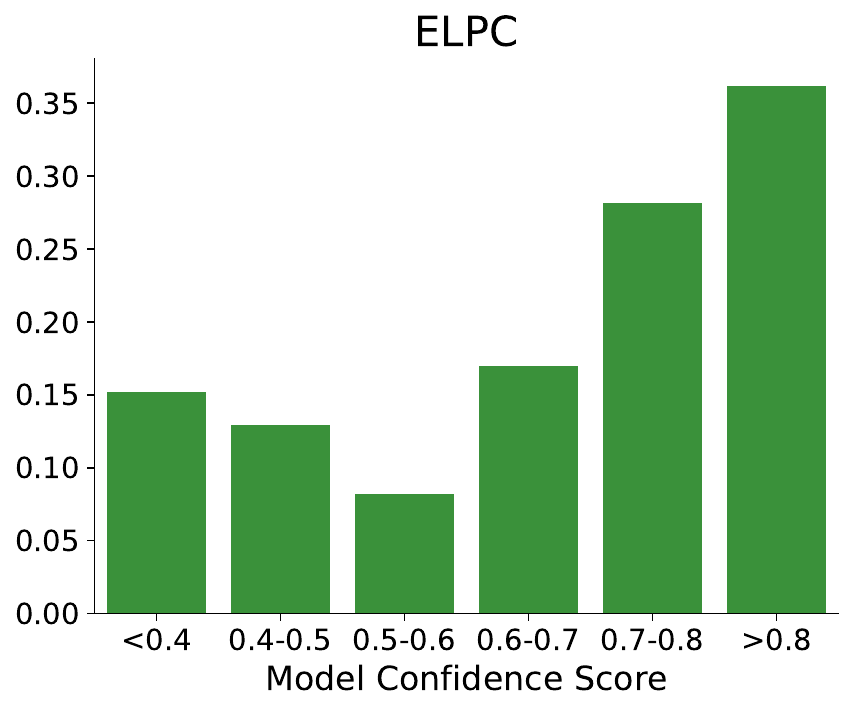}}

\centering
\caption{Detection validation rates by model confidence for WDNR (A, B) and ELPC (C). (A) and (C) show the overall confirmation rate among all detections sent to each organization while (B) shows the confirmation rate only among detections sent to WDNR that passed initial desk screening and were investigated by specialists, illustrating the value of expert review of model outputs.}
  \Description{Three panels show bar graphs of the rate at which detections within a certain range of model confidence scores were verified to be actual manure by the two organizations.}
\label{fig:model_validation}
\end{figure*}

\begin{table}[b]
  \caption{Overall field trial results}
  \label{tab:trial_toplines}
  \begin{tabular}{lcc}
    \toprule
    &ELPC&WDNR\\
    \midrule
    Detections Sent & 536 & 533\\
    Received Field Follow-up & 383 & 123\\
    Field was visible & 284 & --\\
    Manure Present& 93 & 64\\
  \bottomrule
\end{tabular}
\end{table}

ELPC's verifiers were often able to confirm the presence of manure at the detection locations, documenting the process with photos (see Figure~\ref{fig:elpc_photo} for an example of the satellite imagery paired with the verifier's photo confirming the detection). In aggregate, results from both organizations broadly confirmed both that the model was generally capable of detecting manure spreading (Table~\ref{tab:trial_toplines}) and that the model's confidence scores rank-ordered well with ground truth events. Figure~\ref{fig:model_validation} shows the rates at which detections were verified as true spreading events by model confidence bucket for each organization. For WDNR, all detections statewide (but restricted to fields where CAFOs were approved for spreading outside of the winter) with a confidence threshold above 0.5 were sent for follow-up, and detections that were screened out by state staff were assumed to not be spreading events. Figure~\ref{fig:model_validation}(A) shows the verification rate among all detections sent to the agency, indicating a strong relationship between model confidence and the presence of manure. In Figure~\ref{fig:model_validation}(B), the verification rate among only those detections that passed desk review and were sent for on the ground verification is shown. In the constrast between these two plots, the value of human review is clear: the majority of detections that passed the initial screening were found to be positive, across all confidence buckets, while model detections at lower confidence were less likely to be verified overall. For ELPC, detections (Figure~\ref{fig:model_validation}(C)) were not limited to certain fields, with both the geographic distribution and depth in the confidence score reflecting the locations and capacity of the organization's verifiers, which changed somewhat over time. Despite these different selection processes, the rate of confirming the presence of manure conditioned on model confidence was surprisingly similar across the two organizations: the confirmation rate for detections with confidence score above 0.8 was approximately 35\% for both organization, falling to less than 10\% for detections with a score between 0.5--0.6. Although ELPC had slightly higher verification rates for detections with scores under 0.5 (a range not investigated by WDNR), only 87 of the detections receiving field verification were from this part of the score distribution and they were nonetheless verified as actual manure far less frequently than detections with high confidence scores.

Because the trials with the two organizations were being run concurrently, some detections were sent to both for verification, providing an opportunity to more directly assess their agreement. Unfortunately, this overlap was relatively rare: only 57 detections were sent to both organizations, with just 5 detections that received follow-up by both (see Supplementary Table~\ref{tab:sent_both} for the full crosstabs). In 4 of these 5 cases, the organizations agreed on the presence of manure (1 in which both said there was no manure present and 3 in which both confirmed an actual spread), with WDNR confirming an application in the remaining case.\footnote{The ELPC verifier in this case reported the field was entirely snow-covered, perhaps indicating that the application had since been obscured by more recent snowfall.} Both organizations also found similar, but lower, rates of confirmed spreading for events where that organization was the only one to follow up (8 of 24 detections, 33\%, for ELPC; 6 of 14 detections, 43\%, for WDNR). Although the trials were not explicitly designed to assess the degree of organizational agreement and the overlap is too limited to provide a robust measurement, these results provide reassurance that these distinct trials can provide reliable ground truth for our model's detection of dumping events. 

\subsection{Evaluating Clear Violations}
A key motivation for both the development of our winter manure spreading model as well as the timing of the field trials was the Wisconsin statute making manure applications by CAFOs during the months of February and March presumptively in violation of the law. Although the model can only detect the presence or absence of manure from the satellite imagery it operates on, WDNR was particularly interested in the question of whether this spreading constituted a clear \textit{violation} of existing law. Given their regulatory powers, WDNR was also well-suited to assess this question of compliance by directly following up on the detections in a way that ELPC's verifiers could not.

\begin{figure}[b]
\includegraphics[width=0.47\textwidth]{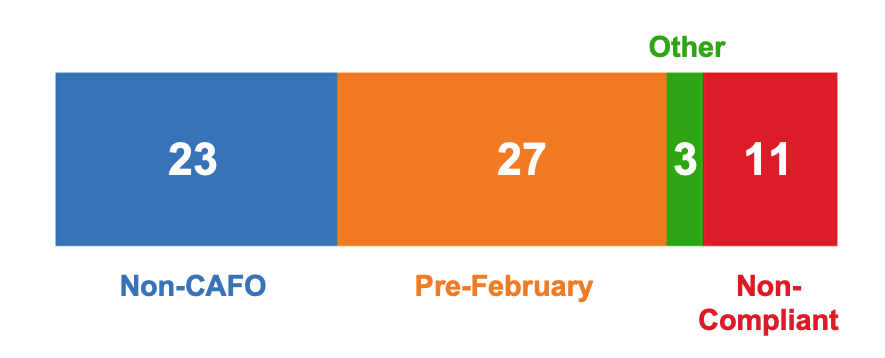} 
\centering
\caption{Determination of regulatory compliance for the 64 spreading events confirmed by WDNR. Only a small fraction (17\%) were found to be non-compliant CAFO spreading.}
  \Description{A bar graph showing the regulatory outcomes for the events verified by WDNR: 11 were non-compliant, 27 were applied prior to February 1, and 23 were applied by smaller farms (3 were determined to be compliant for other reasons).}
\label{fig:wdnr_compliance}
\end{figure}

\begin{figure}[t]
\includegraphics[width=0.4\textwidth]{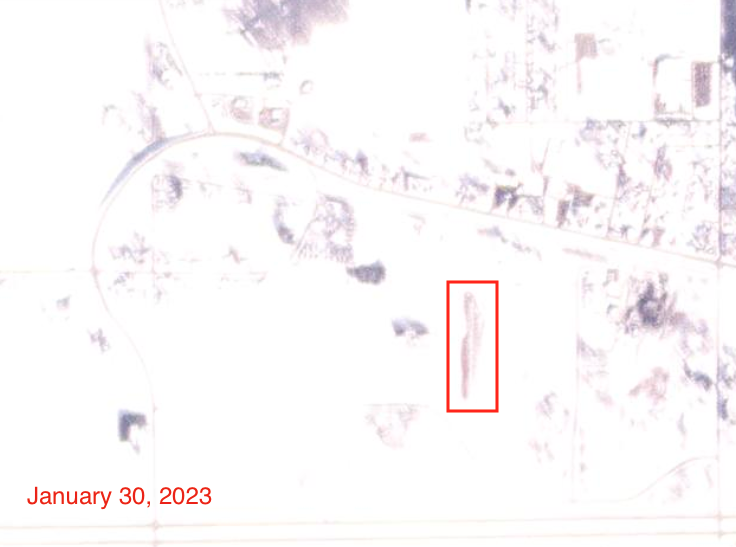} 
\includegraphics[width=0.4\textwidth]{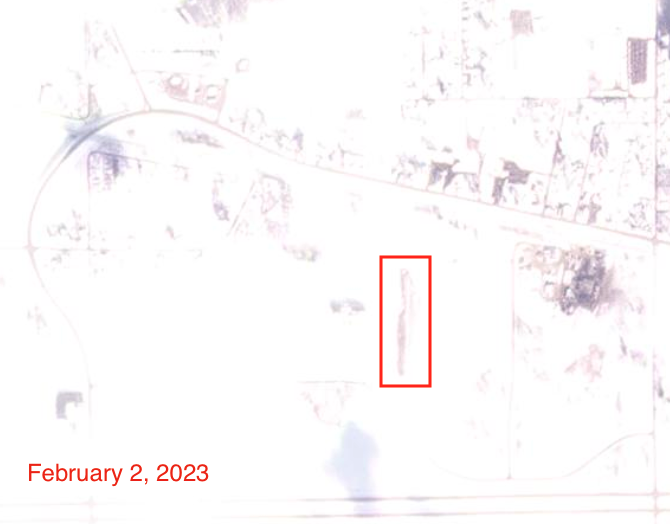} 
\centering
\caption{Example of satellite imagery confirming a manure application detected by our model as of February 2 had been spread in late January, prior to the general restrictions.}
  \Description{The left panel shows satellite imagery captured on January 30, 2023 with a manure application indicated by a red box. The right panel shows imagery of the same location captured on February 2, 2023, with the same application still visible.}
\label{fig:pre_feb}
\end{figure}

Surprisingly, as shown in Figure~\ref{fig:wdnr_compliance}, WDNR staff found very few of the confirmed detections to in fact be clear violations of regulations barring CAFO spreading: Of the 64 events they confirmed as manure spreading events, only 11 were determined to be non-compliant. The remainder were roughly evenly divided among applications that were attributed to smaller AFOs or were reported to have been spread prior to February 1 (and were either still visible or had been re-exposed due to snow melt). In the latter case, the availability of historical satellite imagery allowed us to directly investigate the claim that spreading had taken place earlier in the season. By manually reviewing the imagery at each of these locations, we were in fact able to substantiate many of these claims, finding that at least 18 of the 27 cases had been applied by February 1 (see Figure~\ref{fig:pre_feb} for an example and Supplementary Table~\ref{tab:pre_feb_tab} for full results).

The other explanation for compliant winter spreading --- that the origin was an AFO rather than a CAFO --- cannot be evaluated from the available data as directly. However, our follow-up interviews with WDNR regional specialists detailed the diligence they were taking in verifying these claims. For instance, they described directly reaching out to the smaller farms to confirm that the manure had come from their operations. Their on-the-ground knowledge of the operations of the CAFOs in their jurisdictions was also instructive here, with one specialist noting, ``I went out and saw that it was solid manure and the CAFO in the area only produces liquid manure, so it was easy to determine they didn't apply.'' 
In a follow-up conversation, one of the ELPC verifiers actually concurred with the rarity of CAFO spreading during the winter months, noting that he could not recall encountering occurrences of winter spreading of liquid manure in his many years of living in the area.

Because so many of the confirmed detections were actually determined to be allowable manure applications, we assessed whether compliant and non-compliant detections could be distinguished in satellite imagery. In Supplementary Figure~\ref{fig:comp_vs_non}, we compare the model confidence scores and bounding box sizes of the two types of detections, but find little evidence that non-compliant spreading could be readily distinguished from this imagery. The two populations exhibit very little difference in model confidence and, although bounding box areas of the violating spreads were approximately twice as large on average, the high variance and very small sample size here preclude any robust conclusions, but also call into question the exemption for AFOs based on environmental impact. With only 17\% of all WDNR-confirmed detections determined to be clear violations of current regulations and, even among those applied after February 1, 62\% attributed to smaller, unregulated AFOs (Figure~\ref{fig:wdnr_compliance}), there appear to be substantial risks of environmental harms that fall outside the purview of current regulation.

\subsection{Value of human-ML collaboration}
While ELPC's verifiers sought to follow up on all the detections provided, the pre-screening process employed by WDNR before routing detections for follow-up by regional staff allows us to investigate the value of a human-computer collaborative system in this context. Although it would be prohibitively time-intensive for a human to scan statewide satellite imagery on a daily basis for spreading, reviewing potential events surfaced by a model is far more tractable, and a domain expert may be able to reliably filter false negatives out of these model detections. 
Figure~\ref{fig:model_validation}(B) suggests this is indeed the case with the winter spreading model: although model detections with lower confidence scores are far less likely to be real events at the margin, this relationship disappears entirely after pre-screening. The relatively small fraction of low-confidence detections routed to the field staff were verified at just as high of a rate as high-score detections. Drawing on their expertise and other data sources, the combined human-computer system yielded far higher precision than that of the model alone. Follow-up discussions with our WDNR collaborators indicated that summer imagery was particularly useful here, helping to readily filter out false positive detections that were actually shrubs, trees, buildings, or roadways.

\section{Discussion}
Our field trials broadly confirmed the ability of remote sensing models to detect environmental harms and regulatory violations. Both groups verified the utility of the AI system to spot dumping events, showing how such tools can shed light on dark spaces of environmental compliance, addressing longstanding weaknesses in environmental monitoring arising from its reliance on resource-intensive inspections, self-reporting, or complaints, to detect violations. As the Government Accountability Office reported, ``EPA cannot be certain what its measure [of compliance] is showing and if EPA is making progress toward its goal'' \cite{gao_epa}. Novel data sources, such as satellite imagery and water quality sensors \cite{wei2021attributing}, have substantial promise in providing an independent measure of noncompliance. However, without tools to surface useful and relevant information, this flood of data may overwhelm agency staff already stretched thin. In this project alone, collecting remote sensing data at most twice a week for a single state and during just two months of one year, we reviewed a total of 40,995 images. Manually reviewing this volume of imagery would be intractable, but our computer vision pipeline allowed WDNR to focus only on 533 images that were most likely to show manure application, decreasing the staff work needed to review candidate imagery by 99.8\%. Considered in terms of likelihood that a reviewed image will contain a positive, the model outputs are 77 times more predictive than random selection (12\% vs. 0.16\% positive) for all predictions sent, and 219 times more predictive for the most confident detections. These results illustrate both the potential value of remote sensing data for supporting environmental protection as well as how successfully novel algorithmic tools such as the one deployed here can be used to derive actionable information from these data.

At the same time, our results highlight the importance of organizational context for integrating AI systems effectively. 
The parallel trials with both a nonprofit advocacy group and a government agency uniquely allowed us to explore how the goals and incentives of these two organizations can inform their perspectives on a tool like the winter spreading model. We call this the ``regulatory Rashomon effect'' after the concept (coined for the 1950 Akira Kurosawa film, \textit{Rashomon}) which observes that in complex phenomena, different individuals or groups may give differing or even contradictory interpretations of the same event. Elsewhere, this effect has been documented in ethnography \cite{heider_rashomon}, communication studies \cite{anderson_rashomon_communication}, conservation \cite{Levin_conservation_rashomon} and has even been proposed as a framework for thinking about machine learning \cite{breiman_modeling_cultures}. In this work, WDNR and ELPC confirmed dumping, but exhibited divergent interpretations of the import of the tool. WNDR officials felt that the pilot identified important shortcomings whereas ELPC felt that it demonstrated substantial potential to improve their work to identify environmental harms.

For WDNR, a key consideration in allocating their limited resources is the ability to efficiently detect clear \textit{violations} requiring a regulatory response. From this perspective, the relatively small number of identified violations posed a limitation in terms of the model's utility. Although WDNR staff noted in follow-up conversations that in a typical year only 1-2 violating winter manure applications would be surfaced by the status quo, complaint-driven process (meaning the 11 identified here reflected a 450-1000\% increase over baseline), the process was perceived to entail a significant effort to surface a relatively small number of violations. In general, WDNR staff expressed doubt in the value of implementing the model as part of their current processes without improvements in its ability to distinguish allowable from violating spreading. Improved temporal aggregation methods might be able to filter out pre-February events, but distinguishing AFO vs CAFO applications poses a particular challenge. Nevertheless, WDNR staff noted the pilot provided educational value by starting conversations with CAFOs about winter spreading rules, suggesting that repeating the effort on a 3-5 year cadence could be useful in this regard. Others likewise suggested that awareness of the availability of remote sensing tools to detect these violations might itself provide some deterrence. 
However, they were also cognizant of the negative perceptions some farms might have around the use of AI and satellite data to enforce regulations and the importance of striking a balance between effective enforcement to catch and deter bad actors while maintaining trust with the broader community to foster assistance with good-faith efforts to comply with environmental laws.

As an advocacy organization, ELPC has somewhat broader goals, with interests both in supporting enforcement of existing regulatory regimes as well as advocating for regulatory improvements to better protect the environment. Thus, understanding the extent of both violating CAFO spreading and compliant (but potentially still harmful) spreading by AFOs was important, leading to a more positive outlook on the value of continuing to use the tool going forward. This presents an opportunity for the group to act as a ``force multiplier'' for WDNR's limited capacity by performing initial field screening of model detections to surface confirmed, risky events as complaints. Indeed, through whistleblowing, complaint-driven processes, and citizen suits, advocacy groups often play an important role in supporting regulatory enforcement \cite{defina1997advocacy, zinn2002policing}. Tools like our winter spreading model may improve both the efficiency and comprehensiveness of these efforts. At the same time, the model (combined with field verification) can help the organization raise public awareness of the extent of winter spreading --- whether compliant with current law or not --- and build support for improving regulatory scope and capacity to address the environmental and health risks posed by the practice.

Our results suggest that although a significant amount of winter manure spreading occurs, a sizable portion is applied by smaller farms not regulated under current policy: of the post-February 1 detections confirmed by WDNR, 62\% were determined to be from these smaller operations.
It is also important to note that our work does not attempt to create a population estimate of the full scope of winter manure application. We limited our detection and verification of these events based on resource constraints and priorities of our two partner organizations, focusing on only the most likely to produce positive verification. 
While this design allowed us to pilot the model in conditions that reflect its ongoing deployment (e.g., focusing on organizational efficiency gains), there could be a significant amount of spreading we did not capture, either from AFOs or CAFOs, by not sampling across the full range of model scores. Additionally, our model relies heavily on significant snow cover to contrast the manure applications against. The winter of 2023 was a below average year for snow cover in the state of Wisconsin and we routinely saw weeks where some areas had little to no snow on the ground due to snow melt caused by above average temperatures (see Supplementary Figure \ref{fig:snow_cover_temperature}). Nevertheless, manure application is still prohibited by CAFOs during February and March, regardless of snow cover, and applications could occur on snow and then be melted down to dry earth prior to detection. Further work would therefore be necessary to provide a comprehensive estimate of the full extent of the practice.

Given the considerable fraction of verified applications that were determined to come from smaller farms, we believe this work also has a number of implications for environmental policy, particularly around the use of hard thresholds in regulation given the considerable amount of AFO spreading we observed. The arbitrary 1,000 animal unit threshold raises the question of how well this policy design choice reflects the underlying environmental risks. Such thresholds often elicit ``bunching'' on the favorable side of the line \cite{sneeringer2011effects, houde2022bunching, collins2018bunching}, yet it seems unlikely that the harms from winter manure applications are appreciably lower from a farm with 999 animal units compared to one with 1,001. Because permitting is only required for larger farms in the state, there is little visibility in Wisconsin of the extent to which such bunching might be occurring, and, until this study, no sense of the extent of winter manure spreading from AFOs. However, data from hog farms in Iowa (where farms of all sizes are required to obtain permits) suggests a significant role for this phenomenon in the agricultural space, with a statistically significant large percentage of farms reporting animal counts just shy of this regulatory threshold \cite{sneeringer2011effects}. 

Similarly, the arbitrary February 1 - March 31 prohibition may not be well-tailored to environmental risk. Runoff is determined by snow conditions rather than calendar dates, and historical weather statistics show no indication of a discontinuity in snowfall around the beginning of February (Supplementary Figure~\ref{fig:avg_snowfall}). Such time windows may  also create a perverse incentive to dump in late January to free up storage capacity, which could in fact amplify risks to waterways by clustering spreading temporally. While our preliminary look at the time series of model detections from previous years didn't suggest evidence of significant spreading activity in late January, future work should explore this question more directly.

In the worst case, arbitrary policy thresholds also create an opportunity for less scrupulous actors to evade regulation entirely \cite{giles2022compliance}, a risk exacerbated by the self-reported nature of permitted animal counts.
Remote sensing may also play a role in improving compliance in this regard, for instance by estimating barn size to identify operations that are likely exceeding reported capacities. Even to the extent that thresholds are not being gamed, the operations of moderate-to-large farms shy of the threshold pose concrete environmental and health externalities that suggest a need for better policy designs such as a more graduated approach to regulation.  

Taken together, our dual field trials demonstrate the value AI tools can provide in supporting societal goals like environmental protection, while highlighting the core importance of data science practitioners understanding organizational context in order to successfully deploy such systems and realize this potential. Yet, we also see in our results the promise for these novel technologies to interact with organizational context in a deeper manner. By providing a better accounting of the relative environmental impacts of both regulated and unregulated facilities, this work has helped begin to develop an evidence base for the impact of existing (and seemingly arbitrary) policy design choices that shape the lens through which regulatory agencies such as WDNR evaluate these tools. Our hope, then, is that such tools may also serve to provide a more data-driven approach to ensuring that regulatory structures align well with the underlying social goals they are intended to encode. 


\begin{anonsuppress}
\section{Human Subjects Review}
The human subjects protocol for this work was approved by the Stanford University Institutional Review Board (IRB Protocol \#69822).

\section{Data and Code Availability}
All data associated with model detections sent for validation are available in the Hugging Face data repository reglab/land-app-trial (doi:10.57967/hf/1733), including satellite imagery, metadata about model detections, and results of field validation by both WDNR and ELPC.
%
All code used for obtaining satellite imagery, model inference, and analyzing field trial results is available in a public github repository at github.com/reglab/land-application-detection.
\end{anonsuppress}






\begin{acks}
We thank the Environmental Law Policy Center and Wisconsin Department of Natural Resources, including both the ELPC verifiers and regional WDNR specialists without whose efforts this work could not have been possible. At WDNR, Ben Uvaas and Aaron Orouke were invaluable resources in coordinating our efforts and reviewing imagery. At ELPC, the tireless effort of Andy Olsen, Rob Michaels, Katie Garvey, Brandon Watson, and Pouyan Hatami was essential to co-designing and executing verification efforts. Additionally, we thank Derek Ouyang, Alexandra Haslund-Gourley, and Peter Maldonado for valuable feedback. This work was supported by the Chicago Community Trust and Stanford Impact Labs.
\end{acks}

\bibliographystyle{ACM-Reference-Format}
\bibliography{land_app}

\appendix

\renewcommand{\thetable}{S\arabic{table}}
\renewcommand{\thefigure}{S\arabic{figure}}
\renewcommand{\figurename}{Supplementary Figure}
\renewcommand{\tablename}{Supplementary Table}
\setcounter{figure}{0}
\setcounter{table}{0}

\section{Appendix}

\subsection{Additional Related Work}

\subsubsection{Frontline Regulator Discretion and Incentives}

For the frontline staff tasked with carrying out regulations, the very nature of the role can present conflicting incentives. As one public health official put it \cite{slsblog2018peer}, ``It's easy to forget how independent, and potentially isolating, the job of a food inspector can be. Most of the time is spent doing inspections on your own, engaging with people who are not happy to see you.'' In response, some inspectors may---consciously or not---seek to minimize conflict (or avoid follow-up paperwork) by taking a more lenient approach, and a long history of scholarship has sought to address this challenge. Bardach and Kagan \cite{bardach1982going} describe the unreasonableness of heavy-handed one-size-fits-all regulations that tie the hands of frontline staff and, at worst, may create an atmosphere of mistrust that can undermine a policy's intended goals. Indeed, an over-emphasis on rules and process can produce perverse outcomes, for example, health inspectors in New York must count the number of rodent droppings and score 6 violation points for 30 droppings or 7 points for 31 \cite{nyc2016food}. 

At the same time, a considerable body of work exposes the shortcomings of reserving discretion to the frontline regulators: Paired health inspectors who submitted independent reports were found by one study to disagree 60\% of the time \cite{ho2017does}; Asylum grant rates for refugees can vary by judge from 6\% to 91\%, derided by some authors as ``refugee roulette'' \cite[Figure~22]{ramji2007refugee}; Dramatic variation has been observed in grant rates across patent examiners \cite{lemley2012examiner}; Nuclear Regulatory Commission inspectors find violations at visited reactors with rates ranging from under 10\% to over 60\% \cite[Figure~4]{feinstein1989safety}. Similar levels of inconsistency have been found to plague decisions about Social Security Disability benefits \cite{mashaw1978social}, nursing home inspections \cite{gao2005nursing}, child welfare decisions \cite[Table~3,4]{doyle2007child}, and a host of other policy contexts. Collectively, these failures of regulatory consistency have led to a situation described by Mashaw \cite{mashaw1983bureaucracy} as a breakdown of administrative due process, eroding perceptions of fairness and trust in these core functions of government.

Although the present study does not focus on the variation between individual inspectors, this prior body of work certainly suggests a potential role for the inherent degree of discretion in enforcing environmental regulations to contribute to differences across the organizations we partnered with. If, after all, two health inspectors from the same agency who experience the exact same inspection can disagree more than half the time on violations they observed, certainly we should consider how two individuals working on behalf of separate organizations with their own goals and incentives might perceive the same model results.

\subsubsection{Human-Computer Interaction}

Although the study of organizational behavior and incentives dates back at least a century \cite{weber2019economy}, relatively little work exploring the intersection between human users and machine learning systems has provided a direct empirical investigation of the role of organizational context. Nonetheless, our study builds on an extensive body of work in the human-computer interaction (HCI) literature that touches on various aspects of this question.
For instance, several studies in this space look at AI decision-support systems in various domains including sentencing, healthcare, and child welfare \cite{kostick2022ai, fogliato2022case, sivaraman2023ignore, stevenson2022algorithmic, cheng2022child}. Much of this work has focused on how human decision-makers engage with an AI tool's recommendations, and how the resulting decisions compare to human-only or AI-only systems. Our work contributes to discourse on the effectiveness of humans ``overriding" AI recommendations by observing how staff at the implementing organizations handle model errors through desk review.

While many of these HCI case studies implicitly touch on the role of the organization's broader incentives and structure on the design of the human-AI system, only a handful have explicitly studied these aspects. One such example describes how the process of translating an organization's high-level objectives and goals into a tractable data science problem is often a messy and difficult negotiation between various actors and factors such as data availability \cite{passi2019problem}. Some scholars have written about AI and private sector incentives \cite{slee2020incompatible}, and in particular how Facebook's business model of maximizing engagement has driven the design of its algorithms and platforms \cite{lauer2021facebook, claussen2013effects}. In the public sector, work touching on these aspects has tended to be in the form of higher-level reviews, which draw lessons from multiple examples to highlight the constraints and challenges of implementing algorithms in government \cite{levy2021algoinpublic, veale2019administration, engstrom2023regulating}. These papers raise many important considerations about the role of procurement, data availability, and the lack of capacity for technical model governance in determining how such systems are implemented. 

\subsubsection{HCI and Environmental Monitoring Systems}

In the environmental context specifically, a body of work has focused on how various stakeholders react to the implementation of tech-based monitoring, and what this means for policy outcomes. Case studies of automated monitoring systems have shown how the targets of monitoring can strategically adjust their behavior to either limit the technology's effectiveness \cite{zou2021intermittent}, or fight to prevent its use altogether \cite{browne2023man}. Other work has shown how the self-monitoring power of local governments, who are both subject to federal environmental regulation and are the collectors of pollution monitoring data, has led to cases of strategic shutdowns of monitors preceding anticipated pollution spikes \cite{mu2021strategic}. These studies highlight the need to consider incentives both within and outside the regulatory agency, and explicitly call for more work with external stakeholders to build support for monitoring systems before and during implementation \cite{browne2023man}. 

Our study explores one such model of collaboration, where distinct users of the manure-detection system who have different mandates, exposure to risk, and capacities, use it towards environmental policy goals in different ways. 

\subsection{Supplementary Results}
\subsubsection{Getting the logistics right}

\begin{figure*}[h]
\includegraphics[width=1.00\textwidth]{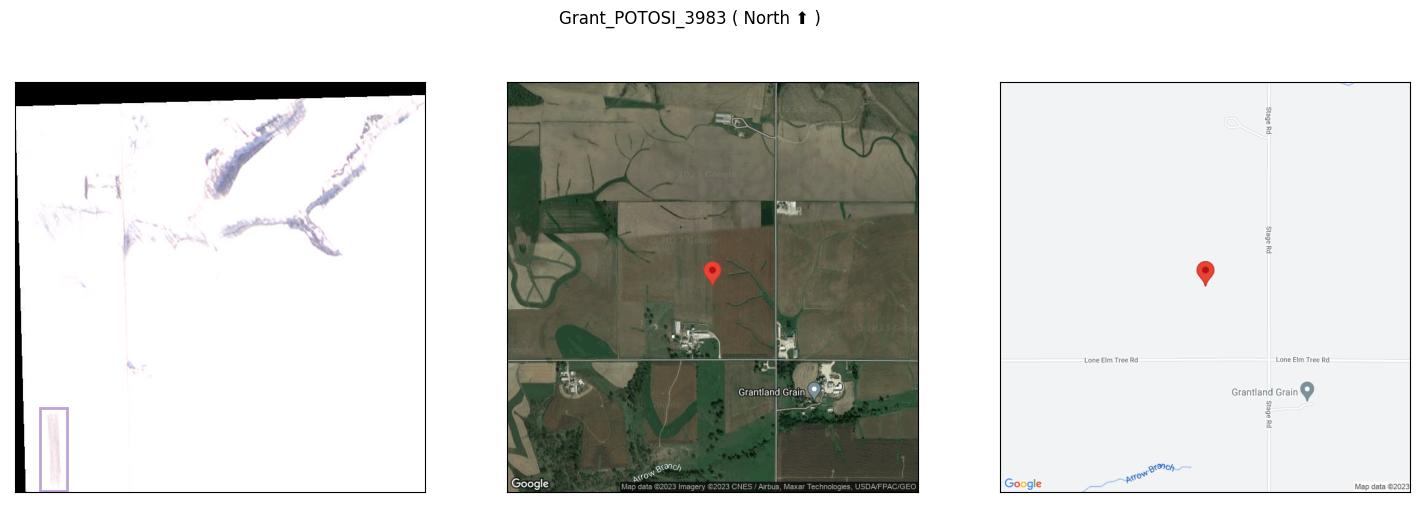} 
\centering
\caption{Example output sent to ELPC verifiers, showing improvements after initial feedback, including a title and indication of north, the addition of a summertime image (middle) better showing nearby landmarks, and a static map image (right) for ease of navigation. The red marker in the middle and right panels indicates the center of the detection bounding box in the left panel (located in the bottom left of the image). Prior to this feedback, only the winter satellite imagery (left panel) was provided.}
\label{fig:elpc-output}
\end{figure*}

Although winter manure application is presumptively non-compliant in Wisconsin only beginning in February, starting the field trial with ELPC's verifiers earlier in the winter allowed us an opportunity to iterate on our processes to improve the data collection during these important months. Frequent check-ins with ELPC surfaced a number of early logistical challenges: for instance, although we were providing links to Google Maps locations of the detections along with satellite imagery, many of these detections were occurring in remote areas with unreliable cellular service, and many of the verifiers were therefore relying on printed materials in the field. Not only did this make the provided Google Maps links less useful, but the often washed-out (or at least snow-covered) nature of the satellite imagery and lack of directional information often meant the orientation got lost as paper copies were shuffled and specified locations were difficult to locate from the ground. To address these problems, we needed to optimize the information provided for the use case of being printed out and brought to the field. Supplementary Figure~\ref{fig:elpc-output} shows an example of the modifications we made, adding a title to the provided imagery, a compass rose to indicate north, a static map image to provide reference for surrounding roads and intersections, and summertime satellite imagery to provide better reference for nearby features that might be visible from the ground.

Because the WDNR trial was more focused on spreading that overlapped with identified CAFO NMP fields that would generally be more familiar with their field staff, the on-the-ground logistical aspects of the work with them were generally simpler. However, initial dry runs were helpful with the state WDNR staff in two ways. First, by reviewing several model detections from the previous winter, they could learn how much effort the pre-screening process would involve and help us determine a model confidence threshold that seemed likely to fit well with their capacity to follow-up on detections and tolerance for reviewing false positives. And, second, this early dry run allowed us to figure out data formats and processes for coordinating on sending detections for review.

Although a pre-screening process was used by state WDNR staff to only route the most likely manure detections for follow-up by CAFO specialists in the field, the ELPC trial sought to develop on-the-ground verification of all model detections that were sent. In doing so, the group's verifiers drove approximately 4,300 miles across the state, spending 175 hours visiting the sites of model detections. Responses were recorded from a similar proportion of detections across the model's confidence score, suggesting our trial results are representative across the range of scores sent (Supplementary Figure~\ref{fig:elpc_process_metrics}A). Likewise, among detection locations visited, verifiers were able to identify the location from public roads 77\% of the time, with a consistent visibility rate across the model score distribution (Supplementary Figure~\ref{fig:elpc_process_metrics}B). These follow-up visits were timely as well (Supplementary Figure~\ref{fig:elpc_process_metrics}C), reducing the risk of manure applications being obscured by thaw or covered by additional snow: among visited sites, 90\% were within 1 day of the detection being sent to ELPC and none took more than 4 days for follow-up.

\subsubsection{Model failure cases}
Both trials also provided valuable information about the types of errors the model could make. Common categories of false positives included rows of hedges, stands of trees, corn stubble, and snow melt exposing cover crops (see Supplementary Figure~\ref{fig:falsepos_example} for examples, including a case where a WDNR specialist noted that recently-installed solar panels were erroneously flagged by the model). In each case, a dark shape offset against a snow-covered background posed a challenge for the model, but the additional human review employed in WDNR's pre-screening indicated that many of these cases could be distinguished and filtered out manually without requiring direct follow-up or a field visit.

Assessing false negatives in a robust manner is much more challenging for a remote sensing project aimed at detecting rare events, but the ELPC trial did provide some insight into these errors as well. In addition to their field reports about model detections, the ELPC verifiers also noted several ``incidental'' manure applications they happened across either while driving to follow-up on a model detection or in the course of their daily lives. In total, they reported 34 such cases: 5 were lacking sufficient information to geocode, 2 were detected by the model but with a confidence score below the threshold for follow-up, 14 were outside of our defined detection area (that is, the 6 km square centered on known CAFO locations), and 13 were in our detection area but not identified by the model. Of these 13, only 1 was visible in manually reviewing the available satellite imagery, with bad quality or missing imagery precluding detection of many of the others. While certainly neither comprehensive nor a representative sample of spreading events missed by the model, these incidental detections do provide some insight into the challenges of both defining an area for investigation (balancing comprehensiveness with the costs of imagery and compute) as well as factors such as cloud cover and poor image quality that can lead to misses by a remote sensing model.

\subsection{Additional Discussion}
Given the extensive literature around frontline discretion and regulatory inconsistency \cite{ho2017does, ramji2007refugee, lemley2012examiner, feinstein1989safety, mashaw1978social, gao2005nursing, doyle2007child}, we might also ask to what extent this mechanism could play a role in the different assessments posed by the two organizations. For instance, could it be the case that a desire by WDNR's regional staff to maintain a good working relationship with the regulated farms or reduce the burden of follow-up and paperwork might result in a tendency to undercount non-compliant spreading? Several lines of evidence here seem to argue against such a mechanism playing an important role in these results: First, the finding that both organizations confirmed the presence of manure at similar rates (as well as agreed with high frequency at sites that were visited by both) suggests it is unlikely that events were being under-verified in general. Second, our ability to independently confirm a large number of the ``pre-February'' events with historical satellite imagery lends considerable credence to this particular alternative explanation. And, third, the diligence of follow-up conducted by WDNR specialists to confirm applications that were attributed to smaller, unregulated AFOs bolstered our confidence in these conclusions. In light of the significant inconsistency that has plagued so much regulatory implementation, this result may seem somewhat surprising. Notably, however, the discrete nature of the assessment here---that is, determining whether manure was present, and, if so, whether it was applied by a CAFO after February 1---is a far cry from the level of nuance and complexity involved in a health inspection or adjudication of disability benefits. Taken together, we feel confident in the conclusion that, although a significant amount of winter manure spreading does take place, a sizable portion of it is applied by smaller farms not regulated under current policy.

\subsection{Supplementary Methods}
\label{method_details}

Attached at the end of the document are:
\begin{itemize}
    \item The script used to guide the semi-structured interviews discussed in the results
    \item The reporting form submitted by ELPC verifiers
\end{itemize}

\clearpage

\clearpage


%

\begin{figure}[h]
\includegraphics[width=0.45\textwidth]{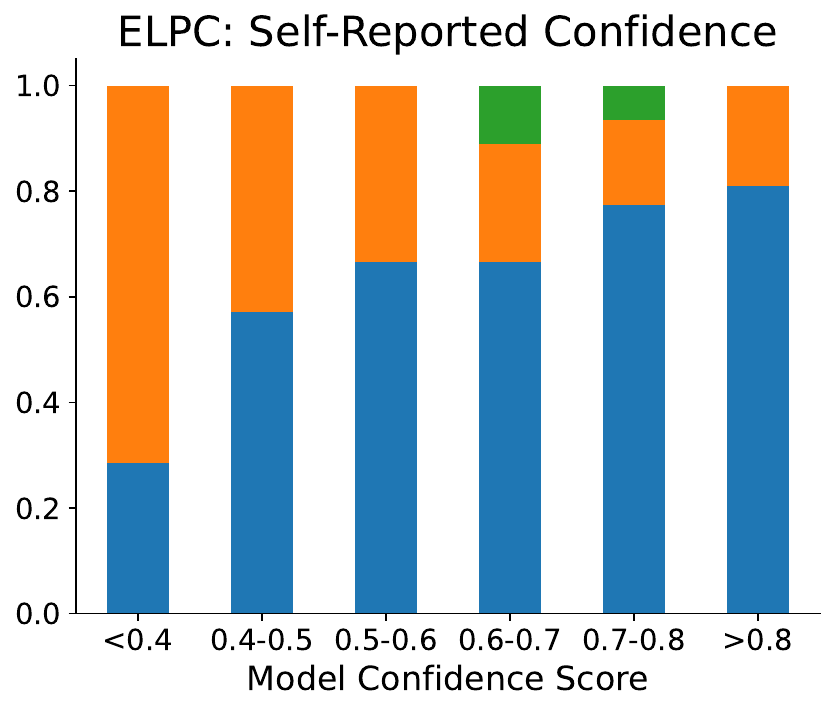} 
\centering
\caption{Self-reported confidence in detection verification by ELPC verifiers generally tracked with model confidence scores. Verifiers were asked to gauge their own level of confidence in their verified detections as high (blue), medium (orange), or low (green)}
\label{fig:elpc_reported_conf}
\end{figure}

\clearpage

\begin{figure*}[h]
\includegraphics[width=0.49\textwidth]{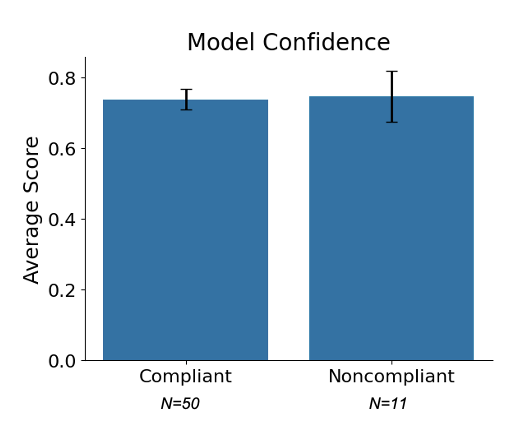} 
\hspace{0.2cm}
\includegraphics[width=0.45\textwidth]{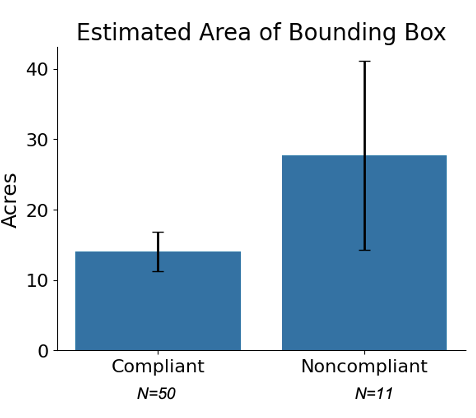} 
\centering
\caption{Compliant and non-compliant spreading events cannot be easily distinguished from model detection results by either average model confidence scores (A) or bounding box areas (B). Error bars are 95\% confidence intervals and 3 events determined to be compliant due to technicalities are excluded.}
\label{fig:comp_vs_non}
\end{figure*}

\clearpage

\begin{figure*}[h]
\includegraphics[width=0.65\textwidth, height=4cm]{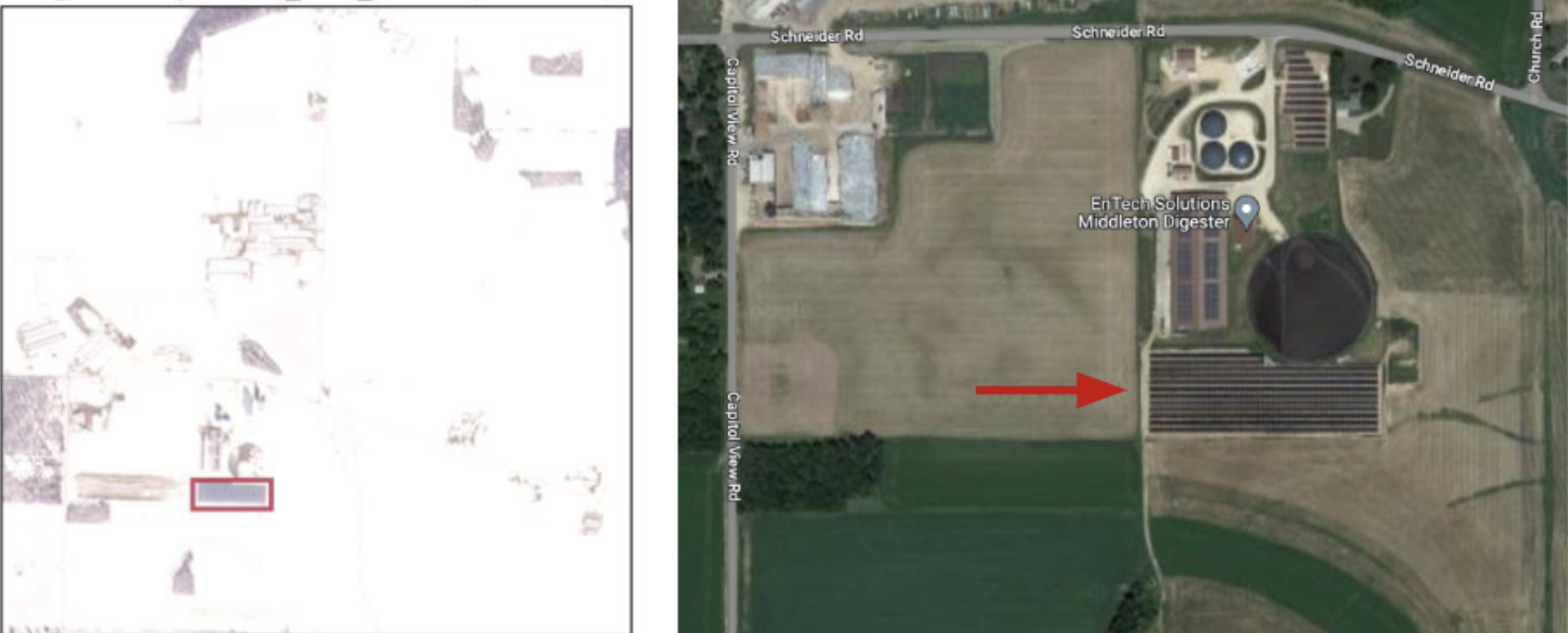} 
\hspace{1cm}
\includegraphics[width=0.267\textwidth, height=4cm]{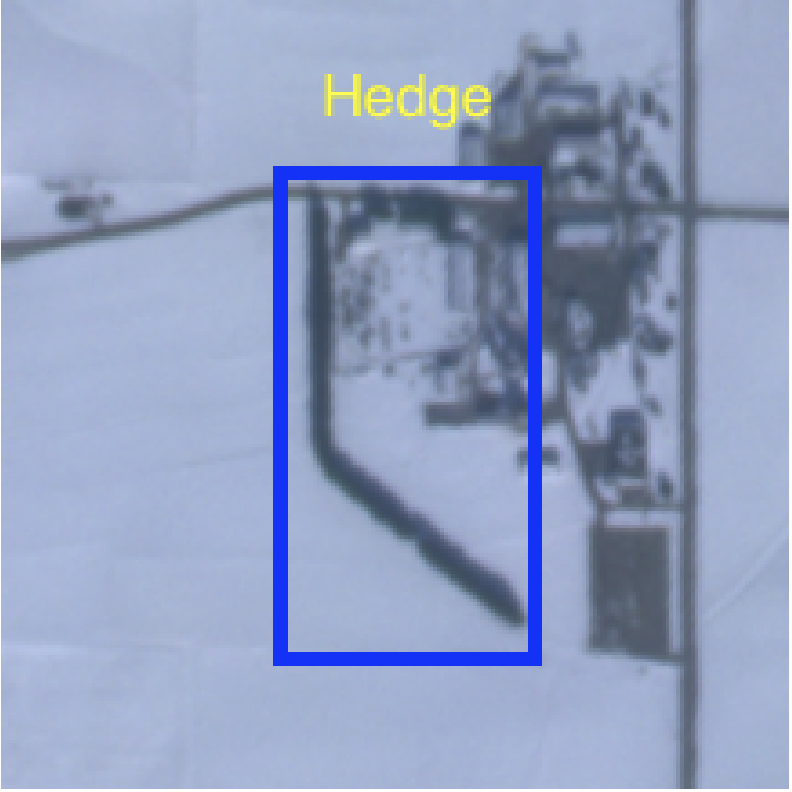} 
\centering
\caption{Examples of false positives detected by the model; the left image is a solar panel, and the right image is a hedge.}
\label{fig:falsepos_example}
\end{figure*}

\clearpage

\begin{figure*}[h]
\includegraphics[width=0.49\textwidth]{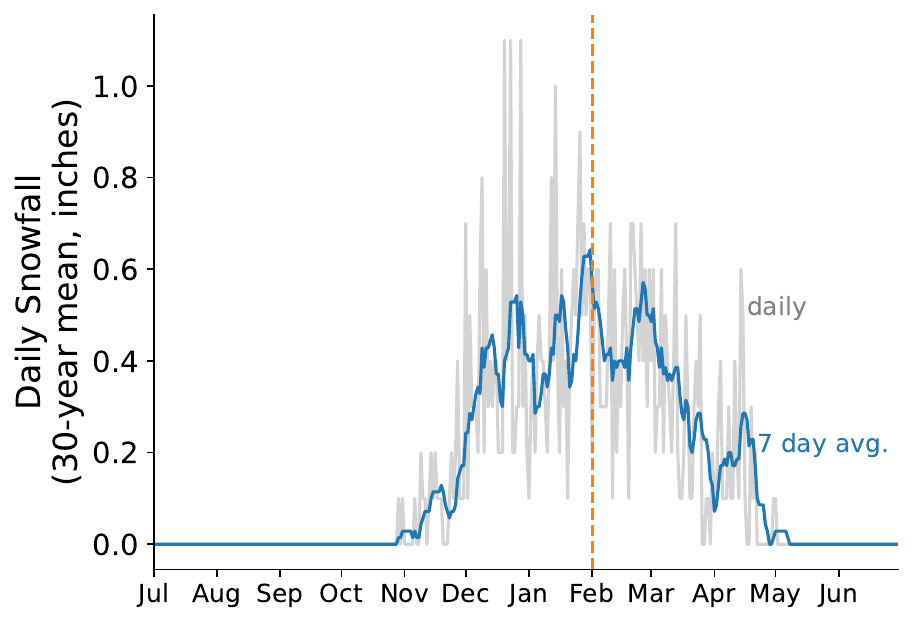} 
\hspace{0.2cm}
\includegraphics[width=0.45\textwidth]{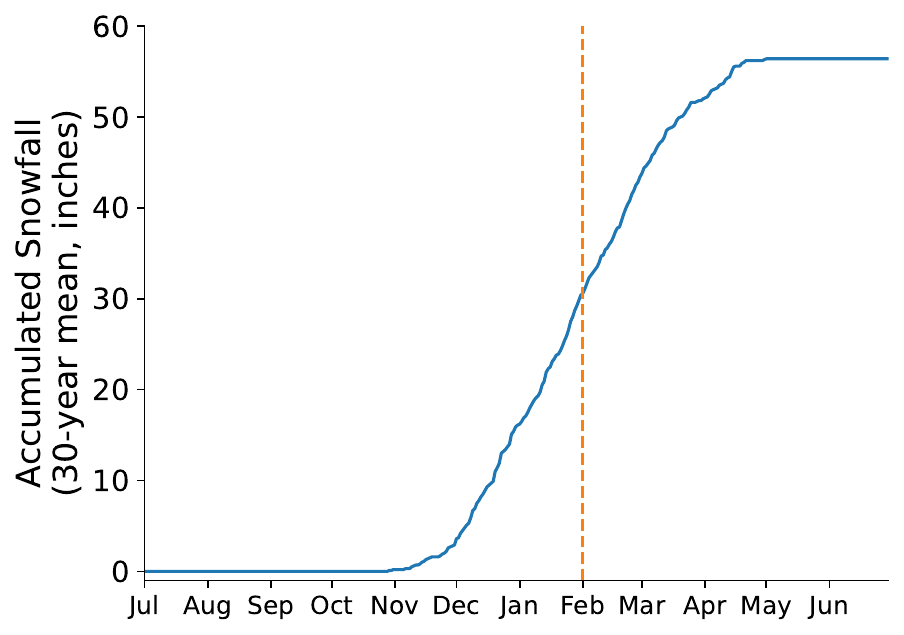} 
\centering
\caption{Historical snowfall statistics in the Green Bay, WI area shows no indication of discontinuity around February 1. (A) 30-year mean inches of snowfall, with raw daily data shown in grey and a 7-day rolling average in blue. (B) Accumulated inches of snowfall, 30-year mean. In both panels, the dashed orange line indicates February 1. All data from US National Weather Service (https://www.weather.gov/wrh/Climate?wfo=grb) with period of record 1993-01-01 to 2023-12-25.}
\label{fig:avg_snowfall}
\end{figure*}

\clearpage

\begin{figure*}
    \includegraphics[width=0.45\textwidth]{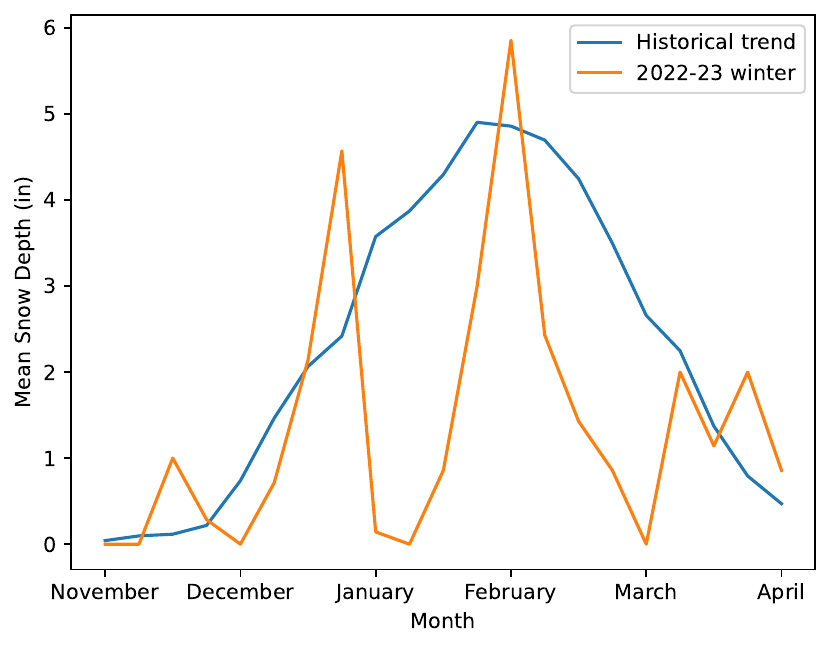}
    \hspace{0.2cm}
    \includegraphics[width=0.45\textwidth]{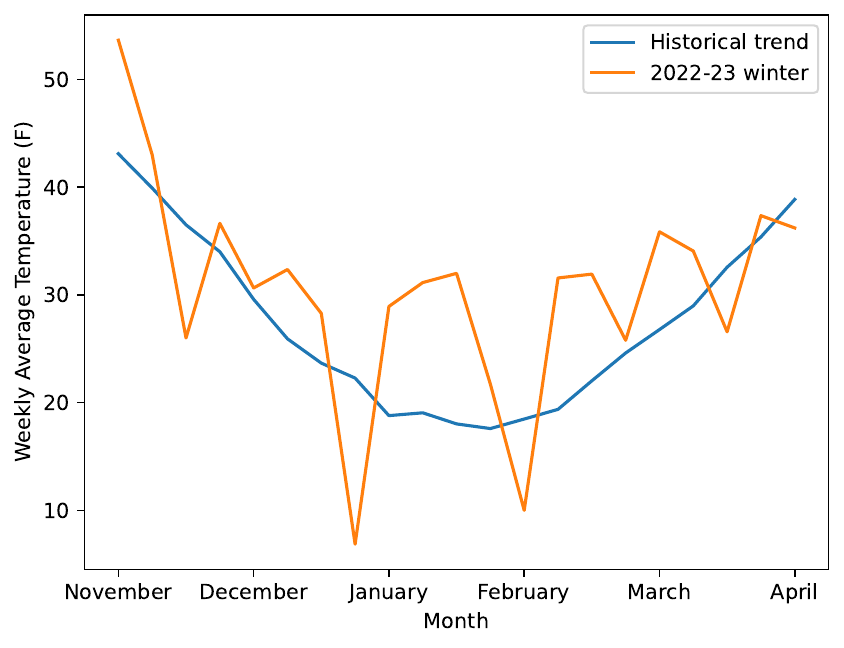}
    \centering
    \caption{Weekly average snowfall and temperature for Madison, WI. Historical mean is for 100 year period ending in 2022. 2023 weekly averages were largely below historical trend for snow cover depth and above trend for temperatures throughout the winter. This resulted in strong melt cycles following snow storms and less snow cover for model inference images.}
    \label{fig:snow_cover_temperature}
\end{figure*}

\clearpage

\begin{table}[h]
  \caption{Verification rates for detections sent to both organizations}
  \label{tab:sent_both}
  \begin{tabular}{rcc}
    \toprule
    &WDNR Follow-Up&WDNR No Follow-up\\
    \midrule
    ELPC Follow-up & 60\% (ELPC); 80\% (WDNR) & 33\% (ELPC)\\
    & \textit{(N=5)} & \textit{(N=24)}\\
    \hline
    ELPC No Follow-up & 43\% (WDNR)& --\\
    & \textit{(N=14)} & \textit{(N=14)}\\
  \bottomrule
\end{tabular}
\end{table}

\clearpage

\begin{table}[h]
  \caption{Manual evaluation of satellite imagery time series for detections determined to be compliant as pre-February spreading}
  \label{tab:pre_feb_tab}
  \begin{tabular}{lr}
    \toprule
    Determination & Number of detections\\
    \midrule
    February Application & 5\\
    Applied Jan 30 to Feb 1 & 7\\
    Pre-February Application & 11\\
    Unsure & 4\\
  \bottomrule
\end{tabular}
\end{table}

\clearpage
\newpage



\section*{Supplementary material}

# Semi-structured Interview Protocol

Hi! My name is _____, and I'm a researcher with [omitted for anonymous review]. Thank you so much for agreeing to be interviewed today. Before we begin, I'd like to share some context about this project and why your participation is so helpful.

First off, thank you for your help so far in our pilot project trying to detect winter manure spreading using satellite imagery. One of our big bets at [omitted for anonymous review] is that the rapidly improving quality and quantity of satellite data has the potential to greatly benefit environmental compliance (beyond manure application, we're also thinking about things like wetlands filling, methane super-emitters, sanitary sewer overflows, etc.).

Our study aims to understand both the opportunities and challenges of putting these new data sources and technologies into practice for environmental regulation, through a combination of quantitative and qualitative methods, including interviews with DNR staff members who have participated in the pilot.

Have you had a chance to read the consent form we emailed to you?

> If no:
> No problem – I will briefly read aloud from a form that describes your rights as a participant. I'll finish by asking you for consent to be interviewed.
> Read consent form verbatim.

Did you have any questions about the form?
If not, or once questions are answered:
Do you give consent to participate in the study? (Verbal YES/NO)
Do you give consent to be audiotaped as part of the study? (Verbal YES/NO)

Wonderful. I want to mention again that this interview is totally confidential. The transcript will not be shared with the DNR or anyone else outside my [omitted for anonymous review] research team, and your name or position won't be tied to anything you share.

Do you have any questions before we begin?

If there are no further questions, hit "Record on this computer" on Zoom

'Set the stage' by sharing relevant background information about the pilot specific to the participant's region or role:

> 

While I have a selection of questions to bring up, I'd really like this to be a conversation - so feel free to speak whatever thoughts or reactions come to mind.

**Part A: Context**
- In a few sentences, could you describe your background in environmental regulation and role with DNR?
  - For how many years have you worked in this area?

If applicable:
- In just a few sentences, could you tell me your role and responsibilities at [facility]?

---

If time:
- Thinking about CAFOs and manure management specifically, how would you characterize the relationship between DNR and the regulated community generally?

---

**Part B: Pilot data and logistics**
- Walk us through your process from getting the detections to following up – generally speaking, what worked well vs not?
- Specifically around the process of following up on a potential detection:
  - How frequently did you end up visiting the location of the application?
  - When you didn't visit, what were the primary other ways you would follow up?
  - What generally drove the decision across these different options (or to not follow up if it didn't seem warranted)?
- In general, how did the regulated farms respond to the pilot?
  - What sort of questions did you get from them? How did you address these?
  - Did you feel like there was more support or information you would have wanted from leadership to respond to these questions and feedback?
- *More questions, perhaps regarding specific detections in this individual's area?*

---

If time:
- Specifically around the format of the information about the detections:
  - What is particularly useful here?
  - Is there additional information you wish you had?
  - Are there changes you might make to the format that would make your job easier?

**Part C: Detection Quality**
- Generally speaking, the model is trying to detect manure spreading from imagery: when you received a detection for follow-up, how frequently would you say it had actually detected manure (whether or not this was a compliant application)?
  - When the model gets it wrong, do you get a sense of what sort of mistakes it's making (for instance, a row of hedges that the model thinks is a spread, etc.)?
- One overall pattern we've found is that a relatively small portion of the detections were in fact determined to be non-compliant applications:
  - In cases where the model detected manure, but it was actually compliant, what does the process look like in terms of making this determination from your end?
  - In those cases, what sort of reasons were most common in explaining why the detected manure spread was compliant?
  - In general, what characterizes non-compliant spreading? How easy or hard do you think it would be to identify those characteristics from satellite images (e.g., on Google Earth)?

---

If time:
- When non-compliant spreading is found, what actions are generally taken (e.g., warnings, notice of violation, etc.)?
- Outside of this pilot, what is the typical process for identifying non-compliant winter spreading like?
  - How does the volume compare in terms of potential leads (e.g., tips, etc), follow-up, and actual non-compliant applications identified?

---

**Part D: Reflection/Cool-down** (~:50-:45)
- Do you have any final thoughts or reflections as we come to the end of the interview?
  - Is there anything I didn't ask that you think is important?

Stop Recording

---

More info to share on the project:
- *[Describe next steps on the project, how they can stay involved]*

Thanks so much for speaking with me today - your perspective is incredibly valuable for this project. Do you have any more questions or concerns?

Thank you again, and have a great day!

Don't forget to send a thank-you note post-interview!

# Winter Manure Spreading Verifier's Report

The name and photo associated with your Google account will be recorded when you upload files and submit this form. Your email is not part of your response.

Any files that are uploaded will be shared outside of the organization they belong to.

*Indicates required question

Email address *

Your answer

Location ID *
(Enter "Incidental" for spreads you came across.)

Your answer

Date Visited *

Date

mm/dd/yyyy

Time of Day *

Time

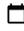 : AM ▾

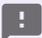

This manure spread: *

○ Was assigned to me

○ I happened to come across it

Is the field visible from public roadways? *

○ Yes

○ No

○ I determined via review of maps. Details below under "Other notes you care to share:"

If the field is not visible, why not? (Distance, obstructions, snow melt)

Your answer

Is farm equipment visible? *

○ Yes

○ No

Description of farm equipment (if applicable):

Your answer

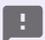

Do you see manure on this field? *

○ Yes

○ No

○ Cannot determine

○ Does not apply

Certainty of observation: *

○ High

○ Medium

○ Low

○ Field not visible

Explain factors affecting the certainty of observation:

Your answer

Is there an odor from the direction of the field? *

○ No odor

○ Detectable odor, but not strong

○ Strong Odor

○ Does not apply

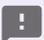

Field is: *

◯ Snow covered

◯ Partially snow covered

◯ Not snow covered

◯ Not visible

◯ Does not apply

Describe predominant weather Conditions: *

☐ Clear

☐ Partially cloudy

☐ Cloudy

☐ Snowy

☐ Rainy

☐ Other: _____

Other notes you care to share:

Your answer

Photos of site:

⬆ Add file

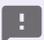

Record your total hours and miles per day, at this [Time Sheets](#) link. This is probably best done when you are filling in your last verifier report of the day.

Your answer

Submit

Clear form

Google Forms

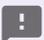



\end{document}